\documentclass{pasj00}
\begin{document}
\SetRunningHead{Y. Osaki}{Helical Tomography for Superhump Light Curves}
\Received{2003/4/2}%{yyyy/mm/dd}
\Accepted{2003/6/12}%{yyyy/mm/dd}

\title{Helical Tomography of an Accretion Disk by Superhump 
Light Curves of the 2001 Outburst of WZ Sagittae}

%%% begin:list of authors
\author{Yoji \textsc{Osaki}}
%  \thanks{Example: Present Address is xxxxxxxxxx}}
\affil{Faculty of Education, Nagasaki University, Nagasaki, 852-8521}
\email{osaki@net.nagasaki-u.ac.jp}

%% `\KeyWords{}' always has to be placed before `\maketitle'.
%Do NOT move this preamble from here!

\maketitle

\begin{abstract}
A new method for analyzing complex superhump light curves for the 2001 
outburst of WZ Sagittae is proposed. The complexity arises 
because intrinsically time-varying and non-axisymmetric distributions 
of superhump light sources are coupled with the aspect effects 
around the binary orbital phase because of its high orbital inclination.  
The new method can disentangle these complexities by separating 
the non-axisymmetric spatial distribution in the disk from the time variation 
with the superhump period. It may be called a helical tomography of 
an accretion disk because it can reconstruct a series of disk images 
(i.e., disk's azimuthal structures) at different superhump phases. 
The power spectral data of superhump light curves of the 2001 outburst 
of WZ Sge by Patterson et al.(2002,PASP,114,721) are now interpreted under 
a new light 
based on the concept of helical tomography, and the azimuthal wave 
numbers of various frequency modes are identified.  
In particular, a frequency component, $n\omega_0-\Omega$, where $\omega_0$ 
and $\Omega$ are the orbital frequency and a low frequency of the apsidal 
precession of the eccentric disk, is understood as an $(n-1)$-armed 
traveling wave in the disk. A vigorous excitation of a wave component 
of $\cos(2\Theta-3\omega_0 t)$ in the first week of the superhump era 
of WZ Sge, where $\Theta$ is the azimuthal angle, supports Lubow's (1991) 
theory of non-linear wave coupling of the eccentric Lindblad resonance
for the superhump phenomenon.  
This method can in principle be applied to other SU UMa 
stars with high orbital inclination if light curves are fully covered 
over the beat cycle.

\end{abstract}

\section{Introduction}
WZ Sagittae, a prototype of one subclass of dwarf novae, called WZ Sge, 
underwent a full-scale outburst in 2001 July and extensive observations, 
particularly in optical light, covering almost all nights without any 
significant gaps, were made for more than one hundred days 
from the start of eruption on July 23 to the long fading to quiescence 
with a worldwide collaboration of amateur and professional astronomers.
Its beautiful light curves can be found in 
VSNET (htpp://www.kusastro.kyoto-u.ac.jp/vsnet.DNe/wzsge01.html) and AAVSO, 
and some of them have already been published by Ishioka et al. (2002) and 
by Patterson et al (2002). The 2001 superoutburst of WZ Sge exhibited many 
interesting features, which will surely be discussed in the future by various 
workers. This paper addresses only one of those features, i.e., an  
analysis of complex light curves of common superhumps observed during 
the main outburst.

WZ Sge exhibited in its photometric light curves periodic humps 
repeating with a binary orbital period called either 
 ``outburst orbital humps" by Patterson et al (2002) or 
``early superhumps" by Ishioka et al (2002) or ``early humps" by Osaki and 
Meyer (2002) for the first twelve days in its 2001 outburst. The origin of 
this phenomenon was discussed by Osaki and Meyer (2002) and by Kato (2002), 
and will not be discussed in this paper. 

After the first twelve days, the common superhumps with a period of 1\% longer 
than the binary orbit emerged in accordance with the SU UMa-type nature 
of WZ Sge stars. The superhumps then subsisted all the way to the 
end of the main outburst and during the so-called echo outbursts 
and in the long fading stage to quiescence (Patterson et al 2002), 
although they may be of ``late superhump nature" in its later stage. 
The overall picture of the 2001 outburst 
of WZ Sge based on the disk instability model (i.e., the thermal--tidal 
instability model) was presented by Osaki and Meyer (2003) 
and may not need to be repeated here.
  
Once superhumps emerge, light curves of WZ Sge become extremely complex, 
as seen,  e.g., in figure 5 of Patterson et al. (2002). Although they look 
extremely complex, one remarkable fact may be recognized, that is, 
the basic patterns of light curves were more or less repeated  
with the beat cycle of about six days between the superhump 
and the orbital periods. The reason why the superhump 
light curves in WZ Sge are so very complex is that intrinsically time-varying 
and non-axisymmetric distributions of superhump light sources are 
coupled with the aspect effects around the binary orbital phases 
because of its high orbital inclination (Osaki, Meyer 2003).  

In this paper we propose a new method of light curve analysis to disentangle 
these complexities by separating the non-axisymmetric spatial distribution of 
the disk from a time variation occurring with the superhump period. 
In section 2, we explain the basic concept of helical tomography 
and how to reconstruct an ``orbital light curve" at a given superhump phase. 
In section 3, we present the inversion method of an ``orbital light curve" 
at a given superhump phase into an azimuthal brightness distribution 
of the disk. 
In section 4 we interpret the power spectra of light curves in the common 
superhump era of the 2001 outburst of WZ Sge observed by 
Patterson et al. (2002) under a new light based on the concept of helical 
tomography, and we identify the azimuthal wave numbers of various frequency 
modes. In section 5 the modes observed in WZ Sge are interpreted in the 
context of Lubow's (1991) theory of non-linear wave interaction 
for the superhump phenomenon. 
Possible applications of the method to other cataclysmic variable stars 
are dealt with in section 6. Section 7 is a summary and discussions.

\section{Basic Concept of Helical Tomography}
It is now well established that the superhump phenomenon is produced 
by a tidally driven eccentric instability within accretion disks 
in cataclysmic variable stars (Whitehurst 1988; Hirose, Osaki 1990; 
Lubow 1991a,b). This instability occurs when an  
accretion disk expands to reach the 3:1 resonance region 
that excites the precessing eccentric structure in the accretion 
disk. 

We first note that the superhump phenomenon is produced by the tidal 
stressing of a precessing eccentric disk by an orbiting secondary star, 
and that the intrinsic variation within the accretion disk 
is strictly periodic with the superhump period. 
We may see how light distributions in the disk 
of the superhump binary system vary with the superhump phase 
in hydrodynamic simulations [e. g., in figure 7b of Hirose 
and Osaki (1990), figure 8 of Murray (1998)]. 
In fact, observed superhump light variations in many 
SU UMa stars (those of low orbital inclination) are singly periodic. 
However, a complication arises when the superhump phenomenon occurs 
in cataclysmic variables with high orbital inclination because aspect 
effects cause other light variations in a non-axisymmetric disk, 
as shown by Simpson, Wood, and Burke (1998).
These complications can be solved by the following method, which we 
call  ``helical tomography".

The method, itself, is very simple in principle. We reproduce 
at the first step an ``orbital light curve" for a certain superhump phase, 
and then invert the ``orbital light curve" thus obtained into 
brightness distribution in the disk for the given superhump phase. 
In this section we explain the method for reconstructing  
an ``orbital light curve" for a given superhump phase. The inversion to 
an azimuthal brightness distribution in the disk is described  
in the next section.

\begin{figure}
\begin{center}
%%BoundingBox: 0 0 548 370
%\FigureFile(19.33cm,13.05cm){helix.eps}
\FigureFile(8.0cm,5.40cm){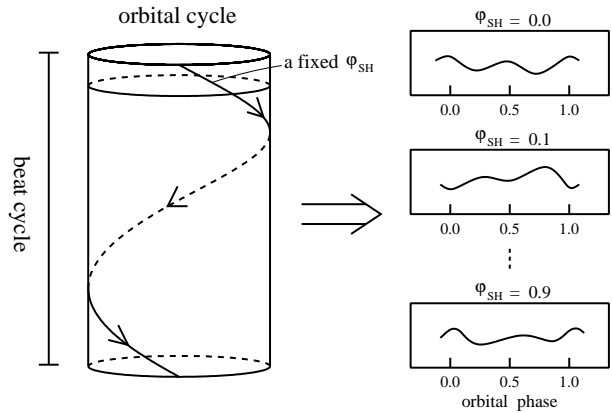}
\end{center}
\caption{%
Schematic illustrations showing how the helical tomography works.
The cylinder in the left stands for light curve observations over 
one superhump beat cycle. The circle of the cylinder corresponds 
to one orbital cycle, while the beat phase proceeds from the 
top to the bottom of the cylinder. 
Let the number of orbital light curves in one beat 
cycle be $N$. The cylinder is then a sum of N thin cylinders.  
Each thin cylinder is thought to show one orbital light curve 
at a certain beat phase, although light curves are not shown here. 
In our helical tomography, we cut a small piece from each orbital light 
curve at a given superhump phase, $\varphi_{\rm SH}$, proceeding 
from the top to the bottom as illustrated by a helix in the left figure. 
We then assemble them into a fictitious ``orbital light curve" 
at a given superhump phase as shown in the right figure. 
In this example we get ten ``orbital 
light curves" for ten superhump phases. Each ``orbital light curve" 
is then inverted to azimuthal light distribution for a given superhump phase 
by a method described in section 3.  
We call this method ``helical tomography" because it can reconstruct 
a series of disk images (i.e., azimuthal light distribution) 
at different superhump phases, and because we cut out pieces of light 
curves ``helically" to construct new light curves for certain superhump 
phases.}
\end{figure}

In order to reconstruct an ``orbital light curve" for a given superhump phase, 
we use all of the superhump light curves for one beat cycle of the orbital 
and superhump periods (i.e.,  about 6 days in the particular case of WZ Sge). 
Figure 1 illustrates the basic process of this method.  Let us consider
an ideal case in which superhump light curves are completely covered 
over the superhump beat cycle, i.e., in the case of WZ Sge, with an orbital 
period of 0.05669 d and a superhump beat period of about 6 d, which results 
in about 100 orbital light curves. We then align these one hundred light 
curves with increasing superhump beat phase, just like figure 5 of Patterson 
et al. (2002). We then cut them into small pieces in which each piece 
corresponds to a certain superhump phase. For instance, let us consider 
to reproduce ten light curves for ten superhump phases, $\varphi_{\rm SH}$. 
We then cut out ten pieces from each light curve, assemble them 
for a certain superhump phase with increasing beat phase, 
and paste them as a function of the orbital phase to construct new light 
curves (in this case ten light curves). 

Each newly formed light curve is a fictitious ``orbital light curve" at a 
given superhump phase, and can be inverted into an azimuthal brightness 
distribution of the disk by a method described in the next section. 
This then gives a disk image at a given superhump phase seen from 
a particular inclination of the binary. 
The advantage of this method is that we can disentangle the geometrical 
effects of the azimuthal light distribution for a given superhump phase 
from the intrinsic time variation of the superhump, itself. In other words, 
we translate the time variations of original light-curves into azimuthal 
light distributions at certain superhump phases. 
We call this method  ``helical tomography" of an accretion disk, 
because it can reconstruct a series of disk images (i.e., 
azimuthal light distributions) at different superhump phases. The ``helical 
tomography" is a medical terminology by which a three-dimensional image of 
a human body is reconstructed. We use this medical terminology 
because the method can be used to reconstruct a series of disk images 
in different superhump phases. The adjective ``helical" is used in another 
sense in that we cut out ``helically" pieces of light curves 
to construct new light curves for certain superhump phases (see, figure 1).    

\section{Inversion of an ``Orbital Light Curve" at a Given Superhump Phase 
into Azimuthal Brightness Distribution of the Disk} 

\subsection {A Simple Model for Superhump Light Distribution} 

In order to extract information about the superhump light distribution 
in the 
accretion disk from an "orbital light curve" at a given superhump phase 
obtained by the method described in the preceding section, we need to make 
some simplifying assumption concerning the superhump light source. 
Radiation from 
the accretion disk of SU UMa-type dwarf novae during the common superhump 
era consists of two different types of radiation. One is radiation from 
the surface of the whole accretion disk produced by the ordinary viscous 
dissipation; the other is that from tidal dissipation near the disk 
edge. When we discuss superhump light curves, it is the common practice 
that we are interested only in time-varying component with a time-scale less 
than the orbital period and we remove from the analysis secular time variation 
with a time-scale longer than the orbital period. In such cases we assume 
that the time-varying part of the radiation is solely produced by the tidal 
dissipation of the eccentric precessing disk, and that radiation 
from the main part of the accretion disk due to the ordinary viscous 
dissipation contributes only to the constant background radiation 
in superhump light curves.       

To extract information about the spatial distribution of superhump 
light at a given superhump phase from  ``orbital 
light curves", we need to make further assumptions. 
Since an orbital light curve includes only ``one-dimensional" information,  
we can extract only one-dimensional information of the disk. Admittedly,  
the terminology ``helical tomography" may be an over-statement in this 
sense because we are not able to reconstruct two-dimensional images 
of the disk. However, we keep using this terminology because we can 
reconstruct a series of the disk images with different superhump phases. 
We adopt a simplest model for radiation from the tidal 
dissipation in a given superhump phase. As discussed by Osaki 
and Meyer (2003),  the tidal dissipation at a given superhump phase 
is non-axisymmetric in its pattern, and occurs near the disk rim 
in a region of about $1/10$ width in the disk radius, i.e., almost 
the same size as the estimated half thickness of the disk at its rim. 
Thus, about one half of the superhump light may be expected to be 
radiated from the two horizontal disk surfaces and one half from 
the vertical disk rim. For binary systems with a high orbital inclination, 
radiation from the tidal dissipation comes mostly from the vertical disk rim; 
its relative importance of the disk rim to the horizontal disk surface 
is proportional to $\tan i$, where $i$ is the orbital inclination. 
Furthermore, we assume for simplicity that 
the surface of the accretion disk is flat, so that radiation 
from the horizontal disk surface for a given superhump phase is 
independent of the orbital aspect. In other words, we may consider 
only superhump light from the vertical disk rim and we may regard radiation 
from the disk surface as the constant background light with the 
above-mentioned assumption when we discuss its aspect dependence 
around the binary orbit.

\subsection{Forward Problem: Transformation of Spatial Distribution of the 
Superhump Light at a Given Superhump Phase into Orbital Light Curve}

To proceed further, we need to make some more simplifying assumptions 
concerning the spatial distribution of the superhump light. 
We here assume that the disk rim is a circular cylindrical surface with 
radius $R$, and vertical width $H$, and that the superhump light 
distribution at the disk rim is now represented by $I=I(\theta)$, where 
$I$ is specific intensity of radiation and  $\theta$  is the azimuthal angle 
of a surface element of the vertical rim in the corotating frame of the 
binary, respectively. Here, the azimuthal angle $\theta$ is measured 
in the direction of the binary orbital motion in our convention.  
That is, all of non-axisymmetric effects of the superhump 
light are assumed to be represented by an azimuthal dependence of 
the intensity of radiation at the circular disk rim. 

This assumption is admittedly very crude as the disk rim is not certainly 
circular, but is elongated during the superhump era. Here, we have restricted 
our attention only to the azimuthal distribution of the superhump light 
by neglecting all other effects. That is the reason why I have adopted 
the above-mentioned assumption. Because of this shortcoming, we must 
be careful not to over-interpret the obtained results. 

We represent the direction to the observer by 
$(i,\theta_0)$, where $i$ is the orbital inclination and 
$\theta_0$ is the azimuthal angle to the observer 
in the corotating frame. We here take into account the limb-darkening 
effect of a surface element, which is written by 
\begin{equation}
I=I_{\rm c} \frac{(1+\beta \cos \vartheta)}{1+\beta}, 
\label{eqn:A1}
\end{equation}
where $I_{\rm c}$ is the intensity of radiation along the surface normal, 
$\vartheta$ is the angle between the surface normal and the direction 
to the observer, and $\beta$ is the limb-darkening coefficient. 
The angle between the surface normal and the direction to the observer, 
$\vartheta$, is expressed by 
\begin{equation}
\cos \vartheta=\sin i \sin \theta',
\label{eqn:A2}
\end{equation}
where $\theta'$ is an azimuthal angle of a disk surface element 
in the observer's frame in which the origin is chosen to the direction 
perpendicular to the observer. The azimuthal angles in these two frames are 
related to each other by   
\begin{equation}
\theta'=\theta-\theta_0+\frac{\pi}{2}.
\label{eqn:A3}
\end{equation}

Let us consider some certain superhump light distribution from the disk rim 
at a given superhump phase, which is represented by 
$I_{\rm c}=I_{\rm c}(\theta)$. 
We may expand it into its Fourier components as 
\begin{equation}
I_{\rm c}=\mathop{\sum}_{m=0}^{m=\infty}  I_m \cos (m\theta+\phi_m), 
\label{eqn:A4}
\end{equation}
where $m=0,1,2,3,....$ are zero and positive integers and $\phi_m$ is a 
constant phase for a given $m$. 

In what follows, we discuss each of Fourier components separately.
The radiation emitted to the observer 
from length $Rd\theta$ of the cylindrical surface is $R H \cos \vartheta 
I_m \cos (m\theta+\phi_m)d\theta$ for the $m$-th Fourier component. 
We now integrate the light contribution for the visible surface of 
the cylinder, for which $\theta'$ varies from $0$  to $\pi$.    
The $m$-th component of radiation received by the observer will then be 
\begin{eqnarray}
\lefteqn{F_m (\theta_0)=\frac{R H I_m/D^2}{1+\beta}}\nonumber \\
& &  \times \int_{0}^{\pi}(1+\beta \cos \vartheta) \cos \vartheta 
\cos (m\theta+\phi_m) d\theta',  \label{eqn:A5}
\end{eqnarray}
where $D$ is the distance of the observer from the star. Substituting 
equations (\ref{eqn:A2}) and (\ref{eqn:A3}) into equation (\ref{eqn:A5}), 
we obtain 
\begin{eqnarray}
\lefteqn{F_m (\theta_0)=\frac{R H I_m/D^2}{1+\beta}}\nonumber \\
& &  \times \int_{0}^{\pi}(\sin i \sin \theta'+\beta \sin^2 i \sin^2 \theta') 
\nonumber \\
& &  \times \cos \left(m\theta'+m\theta_0+\phi_m- \frac{m\pi}{2}\right) 
d\theta'  \label{eqn:A6}.
\end{eqnarray}
After some elementary calculations, we can show that 
equation (\ref{eqn:A6}) can be written as  
\begin{eqnarray}
\lefteqn{F_m (\theta_0)=\frac{R H I_m/D^2}{1+\beta}\sin i }   \nonumber \\
& &  \times  (A_m+\beta \sin i B_m) 
\cos (m\theta_0+\phi_m),
\label{eqn:A7}
\end{eqnarray}
where $A_m$ and $B_m$ are numerical constants. 

Equation (\ref{eqn:A7}) is the very equation which we have sought, 
and it tells us that the orbital light variation due to the $m$-th 
component of 
superhump light is proportional to the intensity at the very surface 
element of the disk rim facing to the observer, i.e., the surface element 
at the azimuthal angle $\theta_0$. The coefficients 
$A_m$ and $B_m$ represents how much reduction of light occurs due to 
integration over the visible surface of the disk rim. As shown below, 
they decrease rapidly by going to the higher order $m$-mode because of 
cancellation effects with positive and negative light contributions. 

 It is found that the coefficient $A_m$ is given for an even integer $m$ by 
\begin{equation}
A_m=(-1)^{1+m/2}\frac{2}{m^2-1}, 
\label{eqn:A8}
\end{equation}
while $A_m=0$ for odd integers $m$, except for $m=1$. The case for $m=1$ is an 
exception and $A_1=\pi/2$. Similarly, the coefficient $B_m$ is given for 
odd integers by 
\begin{equation}
B_m=(-1)^{(m+1)/2}\frac{4}{m(m^2-4)},
\label{eqn:A9}
\end{equation}
while $B_m=0$ for even integer $m$, except for $m=0$ and $m=2$, for which 
$B_0=\pi/2$ and $B_2=\pi/4$.  
Equation (\ref{eqn:A7}) may be written in a more convenient form as 
\begin{eqnarray}
F_m (\theta_0)&=&\frac{2RH}{D^2} \sin i I_m C_m \cos (m\theta_0+\phi_m) \\
\label{eqn:A10}
              &=&F_m^0  \cos (m\theta_0+\phi_m),
\label{eqn:A11}
\end{eqnarray}
where the coefficient 
\begin{equation}
C_m =\frac{A_m+\beta \sin i B_m} {2(1+\beta)} 
\label{eqn:A12}
\end{equation}
is the transformation coefficient from the brightness distribution of the disk 
rim to the orbital light curve. It is normalized to one for a disk rim 
with uniform brightness and no limb darkening. Their  
numerical values are listed in table 1 for two cases with no limb darkening 
and with limb darkening of $\beta=0.56$ and an inclination of $i=90^\circ$ 
together with coefficients $A_m$ and $B_m$. 

\begin{table}
\caption{Coefficients $A_m$ and $B_m$, and coefficient $C_m$ 
for two cases with  no limb darkening and with limb darkening 
of $\beta=0.56$ and $i=90^\circ$.}\label{tab:coeff}
\begin{center}

\begin{tabular}{ccccc}
\hline \hline
$m$&$A_m$&$B_m$ & $C_m$ & $C_m$\\
&&&  ($\beta=0$) &($\beta=0.56$) \\
\hline
0 & 2      & $\pi/2$  &1    & 0.923\\
1 & $\pi/2$  & 4/3    & 0.785 & 0.743\\
2 & 2/3  & $\pi/4$    & 0.333 & 0.355\\
3 & 0 & 4/15 & 0 & 0.048\\
4 & -2/15 & 0 & -0.067 &-0.043\\
5 & 0 & -4/105 & 0 & -0.00068\\
6 & 2/35 & 0 & 0.029 & 0.018\\
\hline
\end{tabular}
\end{center}
\end{table}

\subsection{Inversion: Reconstruction of Azimuthal Brightness Distribution 
of the Disk from Orbital Light Curve} 
The inversion from the orbital light curve to the azimuthal brightness 
distribution is straightforward in our case. Let us write an orbital light 
curve as $F=F(\varphi_0)$ for a certain superhump phase obtained 
by the method described in section 2, where $\varphi_0$ is the orbital 
phase in angular units. The orbital phase $\varphi_0$ and the azimuthal 
angle of the observer in the corotating frame of the binary, 
$\theta_0$, used in the 
previous subsection are related each other by 
$\varphi_0=2\pi-\theta_0$. Since $F=F(\varphi_0)$ is a periodic function 
with $2\pi$, 
we may expand the orbital light curve into Fourier series as 
\begin{eqnarray}
F&=&\mathop{\sum}_{m=0}^{m=\infty}  F_m^0 \cos (m\varphi_0-\phi_m) 
\label{eqn:B1}\\ 
 &=&\mathop{\sum}_{m=0}^{m=\infty}  F_m^0 \cos (m\theta_0+\phi_m), 
\label{eqn:B2}
\end{eqnarray}
where $F_m^0$ is a Fourier amplitude with $m$-th component.  
Here $\phi_m$ is a 
constant phase of the $m$-th component and a minus sign is attached to  
$\phi_m$ in equation (\ref{eqn:B1}) because it has turned out that 
the constant phase $\phi_m$ defined in this subsection is identical 
to that defined in the preceding subsection. We note here that the zeroth 
component, $F_0$, includes not only contributions from the vertical disk rim,  
but also from the horizontal disk surface.    

We now recover individual mode amplitudes, $I_m$, of the azimuthal brightness 
distribution of the disk by inverting equation (\ref{eqn:A10}) to obtain 
\begin{equation}
I_m=\left(\frac {D^2}{2RH\sin i}\right)\frac{F_m^0} {C_m} . 
\label{eqn:B3}
\end{equation}
Once we obtain all Fourier amplitudes, we reconstruct the disk image by 
adding all Fourier components in equation (\ref{eqn:A4}).

\section{New Light on the Power Spectra of Superhump Light Curves 
for the 2001 Outburst of WZ Sge}
Patterson et al. (2002) made a power spectral analysis of 
superhump light curves of the 2001 outburst of WZ Sge over the full 
beat cycle, and obtained many frequencies; all were linear combinations 
of $\omega_0$ and $\Omega$, where $\omega_0$ and $\Omega$ are 
the orbital frequency and the putative apsidal frequency of 
an eccentric disk, respectively. 
As a matter of fact, we do not even need to construct new ``orbital light 
curves" to demonstrate the power of the helical tomography if we already 
have information of power spectra of superhump light curves. 

\begin{table}
\caption{Observed power spectra of superhump light curves 
in the first week of the common superhump era for the 2001 outburst 
of WZ Sge (data from Patterson et al 2002).}\label{tab:obs}
\begin{center}

\begin{tabular}{ccccc}
\hline \hline
$n$&Frequency&Power&Relative amp. &Identification\\
&(c d$^{-1}$)&&$a_n$& \\
\hline
1 & 17.49 & 1030      & 32 & $\omega_0 - \Omega$      \\
2 & 17.64 & $\sim$ 60 & 7.7 & $\omega _0$             \\
\hline
3 & 34.97 & $\sim$ 85 & 9.2 & $2\omega _ 0 - 2\Omega$ \\
4 & 35.12 & $\sim$ 35 & 5.9 & $2\omega _ 0 - \Omega$  \\
\hline
5 & 52.77 & $\sim$ 50 & 7.1 & $3\omega _ 0 - \Omega$  \\
\hline
\end{tabular}
\end{center}
\end{table}

We explain this by using Patterson et al's (2002) data for the first week 
of the common superhump era of the 2001 outburst of WZ Sge. Table 2 summarizes 
results of power spectral analysis taken from their figure 6.  
Each column of table 2 indicates from left to right, the ordinal number, $n$, 
of individual modes with increasing frequency, cyclic frequency in units of 
cycle d$^{-1}$, power of modes estimated by eye from their figure 6, 
relative amplitudes of modes, $a_n$, and its identification. 
As can be seen in their 
figure 6, five discrete peaks in power spectra are seen and there is 
no ambiguity of their identification. 
Two largest peaks at $\omega_0-\Omega=17.49$ c d$^{-1}$  
and $2\omega_0-2\Omega=34.97$ c d$^{-1}$  are those of the superhump period 
and its first harmonics, corresponding to the superhump light curve shown 
by Patterson et al.(2002) at the lowest left frame of their figure 6, 
which was obtained by synchronous summation at the superhump period. The most 
interesting is the signal at $3\omega_0-\Omega=52.77$ c d$^{-1}$  
which appears 
as an isolated peak at the frequency range of the second harmonics 
and its side bands. As will be shown below, it is identified a traveling 
two-armed spiral wave. 

The light intensity in the first week of the common superhump era 
is expressed as  
\begin{eqnarray}
\lefteqn{L(t)}\nonumber \\
&=&L_0+a_1\cos \left[(\omega_0-\Omega)t+\psi_1\right] 
+a_2\cos (\omega_0 t+\psi_2)\nonumber \\
& &+a_3\cos \left[2(\omega_0-\Omega)t+\psi_3\right]
+a_4\cos \left[(2\omega_0-\Omega) t+\psi_4\right] \nonumber \\
& & +a_5\cos \left[(3\omega_0-\Omega) t+\psi_5\right],\label{eqn:C1}
\end{eqnarray}
where $L_0$ stands for the background light and $a_i$ and $\psi_i$ are 
the amplitudes and constant phases for the respective modes. In equation 
(\ref{eqn:C1}) and following mathematical expressions, we understand 
that the frequencies, such as $\omega_0$ and $\Omega$,  
and phases, such as $\varphi_{\rm SH}$  and $\varphi_{\rm beat}$, 
are given in angular units because there may be no confusion, 
for instance, with cyclic frequencies. 
Equation (\ref{eqn:C1}) is rewritten for later convenience as 
\begin{eqnarray}
\lefteqn{L(t)}\nonumber \\
&=&L_0+a_1\cos \left[(\omega_0-\Omega)t+\psi_1\right]\nonumber \\
& &  +a_3\cos \left[2(\omega_0-\Omega)t+\psi_3\right] 
 +a_2\cos (\omega_0 t+\psi_2) \nonumber \\
& & +a_4\cos \left[\omega_0 t+(\omega_0-\Omega) t+\psi_4\right] \nonumber \\
& & +a_5\cos \left[2\omega_0 t+(\omega_0-\Omega) t+\psi_5\right].\label{eqn:C2}
\end{eqnarray}

Let $N$ denote the number of the orbital light curves in the beat cycle, 
which is given by the nearest integer of the ratio $\omega_0/\Omega$. 
A particular time in the $n$-th light curve is then expressed as 
\begin{eqnarray}
\omega_0 t=2\pi (n-1)+\varphi_0, \label{eqn:C3}\\
(\omega_0-\Omega)t=2\pi (n-1)+ \varphi_{\rm SH}, \label{eqn:C4}
\end{eqnarray}
where $\varphi_0$ and $\varphi_{\rm SH}$ are the orbital phase 
and the superhump phase, respectively. They are related to each other 
by $\varphi_0=\varphi_{\rm SH} +\varphi_{\rm beat}$, where 
$\varphi_{\rm beat}=\Omega t$ is the beat phase.
Substituting equations (\ref{eqn:C3}) and (\ref{eqn:C4}) into 
equation (\ref{eqn:C2}), we obtain
\begin{eqnarray}
L&=&L_0+a_1\cos (\varphi_{\rm SH}+\psi_1)
+a_3\cos (2\varphi_{\rm SH}+\psi_3) \nonumber \\ 
&&+a_2\cos (\varphi_0+\psi_2)
+a_4\cos (\varphi_0+\varphi_{\rm SH}+\psi_4) \nonumber \\ 
&&+a_5\cos (2\varphi_0+\varphi_{\rm SH}+\psi_5).\label{eqn:C5}
\end{eqnarray}

Let us now use the concept of helical tomography, in which each 
``orbital light curve" corresponds to a fictitious light curve 
of a snap shot of the accretion disk [see, e.g., ten snap shots of 
the accretion disk in figure 7 of Hirose and Osaki (1990)] viewed from 
different azimuthal directions. We may regard equation (\ref{eqn:C5}) as 
that describing an ``orbital light curve" at a given superhump phase. 
That is, in the above equation $\varphi_{\rm SH}$ is regarded to be  
a fixed constant, while $\varphi_0=2\pi-\theta_0$ is a variable orbital phase, 
representing the azimuthal angle of the line of sight of an observer 
in the co-rotating frame of the binary. 

We may rewrite equation (\ref{eqn:C5}) as 
\begin{eqnarray}
L&=&L_0+a_1\cos (\varphi_{\rm SH}+\psi_1)
+a_3\cos (2\varphi_{\rm SH}+\psi_3) \nonumber \\ 
&&+a_2\cos (\theta_0-\psi_2)
+a_4\cos (\theta_0- \varphi_{\rm SH}-\psi_4) \nonumber \\ 
&&+a_5\cos (2\theta_0-\varphi_{\rm SH}-\psi_5).\label{eqn:C6}
\end{eqnarray}
By using the method described in the preceding section, we now 
invert equation (\ref{eqn:C6}) to obtain the surface brightness distribution 
of the vertical disk rim as 
\begin{eqnarray}
\lefteqn{I(\theta,\varphi_{\rm SH})\left(\frac {2RH\sin i}{D^2}\right)}
\nonumber \\ 
&=&L_0+b_1\cos (\varphi_{\rm SH}+\psi_1)
+b_3\cos (2\varphi_{\rm SH}+\psi_3) \nonumber \\ 
&&+b_2\cos (\theta-\psi_2)+b_4\cos (\theta- \varphi_{\rm SH}-\psi_4)  
\nonumber \\ &&+b_5\cos (2\theta-\varphi_{\rm SH}-\psi_5),\label{eqn:C7}
\end{eqnarray}
where $b_1=a_1$, $b_3=a_3$, $b_2=a_2/C_1$, $b_4=a_4/C_1$, and $b_5=a_5/C_2$
in the case of no limb darkening. 

Equation (\ref{eqn:C7}) represents the azimuthal brightness distribution of 
the superhump light at the disk rim for a given superhump phase. 
In order to reconstruct the brightness distribution of the disk 
for a certain superhump phase, we need information about the constant phases, 
$\psi_n$, of individual Fourier components. Unfortunately, this information  
is usually not available in published form. 
Since the phase information of individual modes has turned out to be very 
important from the standpoint of the helical tomography, we urge 
observers to publish such information together with the amplitudes of 
individual modes. 

Since $\varphi_{\rm SH}=(\omega_0-\Omega)t$, we may rewrite the above 
equation as 
\begin{eqnarray}
\lefteqn{I(\theta,t)\left(\frac {2RH\sin i}{D^2}\right)}
\nonumber \\ 
&=&L_0+b_1\cos \left[(\omega_0-\Omega)t+\psi_1\right]
+b_3\cos \left[2(\omega_0-\Omega)t+\psi_3\right] \nonumber \\ 
&&+b_2\cos (\theta-\psi_2)\nonumber \\ 
&&+b_4\cos \left[\theta- (\omega_0-\Omega)t-\psi_4\right] \nonumber \\ 
&&+b_5\cos \left[2\theta-(\omega_0-\Omega)t-\psi_5\right].\label{eqn:C8}
\end{eqnarray}
Equation (\ref{eqn:C8}) describes the brightness distribution of the superhump 
light as a function of the azimuthal angle, $\theta$, in the disk 
and time, $t$. As expected, it is a periodic function with the superhump period.

The first line of the right hand side of equation (\ref{eqn:C8})
is then a certain constant at a given superhump phase, 
representing the light level of a synchronously summed superhump light curve 
(see, the lowest left panel of figure 6 of Patterson et al. 2002). 
The second line of equation (\ref{eqn:C8})  represents 
a one-armed wave fixed in the corotating frame of the binary. 
It corresponds to the lowest right panel of the same figure; 
however 
a much clearer figure is found in the lowest right panel of figure 7 
for the second week of the common superhump era (Patterson et al. 2002). 
This was called ``mean orbital light curve", and was interpreted as 
a hot spot light curve by Patterson et al. (2002). As discussed by Osaki 
and Meyer (2003), we have argued that the ``mean orbital light curve" simply 
represents the non-axisymmetric distribution of the superhump light, 
and that it is not due to the enhanced hot spot. This point will 
be taken up once more in the next section. 

The third line also represents a one-armed wave, but this time 
it is a traveling wave which rotates progradely once per superhump 
period in the co-rotating frame, because 
\begin{equation}
\frac{d\theta}{dt}=\omega_0-\Omega. 
\end{equation} 
If we introduce the azimuthal angle in the inertial 
frame by $\Theta$, it is related to that in the corotating frame by 
$\Theta=\theta+\omega_0 t$. Thus, this wave mode rotates twice in the inertial 
frame per superhump period.   

From the same discussion as given above, we find 
that the last term of the right-hand side represents a two-armed traveling 
wave, as  
\begin{equation}
\frac{d\theta}{dt}=\frac{\omega_0-\Omega}{2}. 
\end{equation} 
That is, the wave pattern rotates one half per superhump 
period in the corotating frame. However, the same disk configuration returns 
back after one superhump period because the basic pattern is two-armed. 

%%%%%%%%%%%%%%%%%%%%%%%%%%%%%%%%%%%%%%%%%%%%%%%%%%%%%%%%%%%%%%%%%%%%%%%%%%%%%%%
%begin revised parts(June 11,2003)
The brightness distribution of the disk can also be expressed in the inertial 
frame (i.e., in the observer's frame). Since the azimuthal angle in the inertial frame, $\Theta$, is related to that in the corotating frame, $\theta$, 
by $\theta=\Theta-\omega_0 t$, equation (\ref{eqn:C8}) is rewritten as 
\begin{eqnarray}
\lefteqn{I(\Theta,t)\left(\frac {2RH\sin i}{D^2}\right)}
\nonumber \\ 
&=&L_0+b_1\cos \left[-(\omega_0-\Omega)t-\psi_1\right]
+b_3\cos \left[-2(\omega_0-\Omega)t-\psi_3\right] \nonumber \\ 
&&+b_2\cos (\Theta-\omega_0 t-\psi_2)\nonumber \\ 
&&+b_4\cos \left[\Theta- (2\omega_0-\Omega)t-\psi_4\right] \nonumber \\ 
&&+b_5\cos \left[2\Theta-(3\omega_0-\Omega)t-\psi_5\right].\label{eqn:C9}
\end{eqnarray}
Equation (\ref{eqn:C9}) describes the brightness distribution of the disk 
in the inertial frame. 

The above results are summarized in table 3 where wave modes 
are specified by $(m, \ell)$ in which the brightness distribution in the disk 
is written as  $\cos (m\Theta-\ell \omega_0 t)$. Here, a small frequency 
shift of $\Omega$ to $\ell \omega_0$ is disregarded in specifying 
the wave modes, 
because  $\Omega\ll \omega_0$. The mode identification of five modes  
with $(m, \ell)$ is straight forward from equation (\ref{eqn:C9}). 
The main peak of the power spectrum (i.e., the $n=1$ mode) is 
the well-known superhump periodicity with frequency $\omega_0-\Omega$ 
and its wave is identified by the $(0,1)$-mode 
because there is no dependence on the azimuthal angle, $\Theta$, 
as seen from equation (\ref{eqn:C9}). The slowly precessing eccentric disk 
mode with $\cos (\Theta-\Omega t)$-wave is identified as the $(1,0)$-mode 
but this mode, itself, does not produce any brightness distribution 
except for secular variation with the beat period (i.e., $2\pi/\Omega$). 
The tidal coupling of the eccentric $(1,0)$-mode with the companion's 
gravitational field described by $(m,m)$-waves produces $(m-1,m)$-modes 
as discussed in Lubow (1991a) and in the next section. The superhump light 
variation with the $(0,1)$-mode is produced by a tidal interaction between 
the eccentric $(1,0)$-mode and the tidal $(1,1)$-mode. The various wave 
modes excited in the superhump disk are discussed in the next section.

%end revised parts(June 11,2003)
%%%%%%%%%%%%%%%%%%%%%%%%%%%%%%%%%%%%%%%%%%%%%%%%%%%%%%%%%%%%%%%%%%%%%%%%

\begin{table}
\caption{Mode identifications and inverted amplitudes of 
individual modes.}\label{tab:inversion}
\begin{center}

\begin{tabular}{ccccc}
\hline \hline
$n$ & Frequency& Identification&Mode&Inverted amp. \\
&(c d$^{-1}$)&&$(m,\ell)$ & $b_n$ \\
\hline
1 &17.49 & $\omega_0 - \Omega$      & (0,1)& 32 \\
2 & 17.64 & $\omega _0$              & (1,1)& 9.8 \\
\hline
3 & 34.97 & $2\omega _ 0 - 2\Omega$ & (0,2)& 9.2 \\
4 & 35.12 & $2\omega _ 0 - \Omega$  & (1,2)& 7.5 \\
\hline
5 & 52.77 & $3\omega _ 0 - \Omega$  & (2,3)& 22 \\
\hline
\end{tabular}
\end{center}
\end{table}

We find from table 3 that the wave mode with $\cos (2\Theta-3\omega_0 t)$ 
was very strongly excited in the first week of the common superhump era 
in the 2001 outburst of WZ Sge; its significance is discussed 
in the next section.

The above result can easily be extended to a Fourier component with frequency 
$n\omega_0-\Omega$. Since the time dependence of this mode 
is written as 
\begin{equation}
\cos \left[(n\omega_0-\Omega) t+\psi_{n}\right] 
=\cos \left[(n-1)\varphi_0+\varphi_{\rm SH}+\psi_{n})\right],
\end{equation}
it represents an $(n-1)$-armed traveling wave which rotates 
$\left[1/(n-1)\right]$-times 
per superhump period. However, the basic disk configuration returns  
back after one superhump period because of its $(n-1)$-armed structure.
Patterson et al. (2002) found in the second week of the common superhump era 
of the 2001 outburst of WZ Sge the primary Fourier components 
at $n\omega_0-\Omega$ in the power spectrum 
with $n=1, 2, 3, 4, 5, 6, 7, 8, 9$. These are exactly $(n-1)$-armed waves 
discussed here. The appearance of higher Fourier components in the second 
week indicates the appearance of a much shaper feature in the second week 
for the azimuthal brightness distribution in the disk.

\section{Lubow's Theory of Non-Linear Wave Coupling by 
the Eccentric Lindblad Resonance}

In the previous section we have demonstrated that by using the concept of 
helical tomography we can identify azimuthal wave numbers of various 
frequencies found in the power spectral analysis of superhump light 
curves by Patterson et al (2002). 
In particular, we identify the Fourier components at $n\omega_0-\Omega$ as 
$(n-1)$-armed traveling waves. In this section, we discuss why these wave 
modes are excited in the accretion disk of a superhump binary. 
To do so, we use Lubow's theory (1991a,b) of non-linear wave coupling. 
Below, we basically follow Lubow's (1991a,b) discussions on this problem.

In this section, we use mostly the inertial frame of reference rather than 
the corotating frame. The tidal perturbing potential in the accretion disk 
by the secondary star is expressed as 
\begin{equation}
\phi(r,\Theta)=\phi_m (r) \cos(m\theta)=\phi_m (r) 
\cos\left[m(\Theta-\omega_0 t)\right].
\end{equation}   
Let us consider wave fields in the accretion disk which are specified 
by two integers $(m,\ell)$  in which its azimuthal and time dependence is 
expressed as $\cos \left(m\Theta-\ell \omega_0 t\right)$. 
The tidal perturbation potential produces 
forced wave fields with $m=m$ and $\ell=m$, as is known very well. 

Let us now impose some finite eccentricity in the accretion disk. The disk 
eccentricity is described as a one-armed eigenmode whose eigenfrequency, 
$\Omega$, is known to be very small as compared with $\omega_0$ 
(see, Hirose, Osaki 1993). Following Lubow (1991a), we denote this mode 
as the (1,0)-mode. The non-linear wave coupling between the $(1,0)$-mode and 
the $(m,m)$-mode produces another wave mode with $(m-1,m)$. Lubow (1991a,b) 
has demonstrated that, in particular, a strong wave with $(2,3)$ is 
launched at the 3:1 resonance region of the accretion disk by the 
eccentric Lindblad resonance. Further coupling between 
the $(m-1, m)$-mode and the $(m,m)$-mode reinforces the imposed eccentric 
mode with (1,0). This is Lubow's picture of excitation of the eccentric mode 
by the eccentric Lindblad resonance.    

So far we have neglected the small, but finite, frequency, $\Omega$, of 
the eccentric mode to describe various $(m,\ell)$-modes. Let us now 
take into account slow precession of an eccentric mode (1,0). This 
mode has a small, but finite, frequency of $\Omega$, describing a beat period 
of about 6 days in the case of WZ Sge. Since the $(m-1,m)$-mode is produced 
by coupling between the (1,0)-mode and the $(m,m)$-mode, its frequency is 
given by $m\omega_0-\Omega$. This is exactly the $(m-1)$-armed traveling 
wave discussed in the previous section. 

The $(0,1)$-mode with frequency $\omega_0-\Omega$ corresponds to 
the main peak of the power spectrum of the 
common superhump era with the superhump period, only a periodicity 
to be observed in superhump binary systems with a low orbital inclination.   
Many peaks in the power spectra found by Patterson et al (2002) 
in the common superhump era are those with $m\omega_0-\Omega$, which are 
interpreted as the $(m-1,m)$ mode in the previous section. The dominance of 
these modes in the common superhump era is a natural consequence of Lubow's 
(1991) theory. In particular, the fact that the $(2,3)$ mode at a frequency 
$3\omega_0-\Omega$ is very strong [see, the right most panel of the middle 
row in fig. 6 of Patterson et al. (2002)] in the first week of 
the common superhump era, supports Lubow's theory because the excitation 
of the (2,3)-mode plays the most important role for the growth of 
the eccentric mode by the 3:1 resonance in his theory. 
The (2,3)-mode is two-armed spiral waves, rotating one and half times 
around the accretion disk over one superhump period in the inertial 
frame, as clearly seen in hydrodynamic simulations by Simpson and Wood (1998).  

Another interesting aspect is a strong excitation of a mode 
with frequency $\omega_0$, in particular, in the second week of 
the common superhump era of the 2001 outburst of WZ Sge. 
This is clearly seen in the lowest right panel of figure 7 by 
Patterson et al. (2002), and it was called the ``orbital 
light curve" because it was obtained by synchronous summation  
at the orbital period. Patterson et al (2002) interpreted it 
as being a hot-spot 
light curve, and they used it as evidence for enhanced mass transfer. 
This mode is a one-armed wave pattern with $(1,1)$ fixed in the 
corotating frame of the binary, as discussed in the previous section. 

In Lubow's theory for the excitation of eccentricity of a disk, 
it is a two-step process. In the first stage, the (1,0)-mode grows 
rapidly together with the (2,3)-mode launched by the 3:1 resonance.
In the second stage, once the (1,0)-mode attains a large amplitude, 
various non-linear wave couplings become important, and a kind of 
steady state is established between various wave modes. 
In this picture, the (1,1)-mode is also reinforced by non-linear 
wave coupling between the original eccentric (1,0)-mode 
and the (0,1)-mode responsible for the superhump light variation. 
The (1,1)-mode observed with a large amplitude in the second week 
is thus a result of superhump phenomenon itself. If so, the amplitude 
of the (1,1)-mode will vary in a long run together with the amplitude 
of the (0,1)-mode; in other words, the amplitude of ``orbital wave" 
in Patterson et al. (2001) 
should vary together with the superhump light amplitude, a picture already 
presented by Osaki and Meyer (2003).  

\section{Application to Other Systems}
An SU UMa-type dwarf nova, IY UMa, is another eclipsing binary system 
for which Patterson et al. (2000) found a rich spectrum of frequencies 
in the power spectral analysis during its superoutburst. Its power 
spectrum is very similar to that of WZ Sge, because signals 
at $m\omega_0-\Omega$ with $m=3,4, 5$ were found besides 
the well-known superhump periodicity at $\omega_0-\Omega$. 
They are identified as $(m-1)$-armed traveling waves, just in the 
same way as in the case of WZ Sge, as discussed in sections 4 and 5. 

There exist several eclipsing SU UMa stars.  The most well known among them 
are OY Car, Z Cha, and HT Cas. Their superhump light curves could be examined 
by new light based on the concept of helical tomography.   

The AM CVn stars are thought to be cataclysmic variable stars having 
helium-rich and hydrogen-deficient accretion disks (Warner 1995). 
In particular, its prototype star, AM CVn, itself exhibits very complex 
light variations. As already pointed out by Solheim and Provencal (1998), 
the disk structure will be analyzed by examining a rich spectrum of 
light curves of AM CVn stars. 
By analyzing its light curve, Skillman et al. (1999) have 
obtained periodic signals as many as 20 discrete frequencies. 
According to Skillman et al. (1999), the basic clocks of this star are  
an orbital period of 1028 s and a superhump period of 1051 s. 
However, many of the discovered frequencies can be understood 
as harmonics and sidebands of these fundamental frequencies, i.e., 
with $n\omega_0-m\Omega$, where 
$\omega_0$ and $\Omega$ are the orbital frequency and the apsidal precession 
frequency of an eccentric disk, respectively. On the other hand,  
Solheim et al. (1998) have made a quite different interpretation in that 
the fundamental frequency at $951 \mu$Hz (with a period of 1051s) is related 
to the orbital period of the binary and the dominant light signal at 525 s is 
due to the two-armed spiral structure in the disk fixed in the corotating 
frame of the binary.  

\section{Summary and Discussions}
We have proposed a new method to analyze complex superhump light curves of 
SU UMa stars with high orbital inclination, particularly   
for the 2001 outburst of WZ Sge. It is called the helical tomography because 
it can reconstruct a series of accretion disk images in various 
superhump phases. Based on the concept of helical tomography, 
power spectral data of superhump light curves obtained for the 2001 outburst 
of WZ Sge by Patterson et al. (2002) have been interpreted, and the azimuthal 
wave number of various frequency modes identified. 

As already discussed in section 3, a very crude assumption had to be made 
for the superhump light distribution in order to perform an inversion of 
the orbital light curves into the azimuthal distributions of 
the superhump light. 
This is one of the biggest shortcomings in our theory and a much better method 
should be explored. 
Nevertheless, we can demonstrate that a two-armed traveling wave with 
$\cos (2\Theta-3\omega_0 t)$ was excited vigorously in the first week 
of the common superhump era, supporting Lubow's (1991) theory 
of the eccentric Lindblad resonance for the tidal instability. 

\vskip 3mm

I would like to thank Dr. F. Meyer of Max-Planck-Institut f\"ur Astrophysik 
and Dr. M. Takata of The University of Tokyo for helpful discussions. 
This work is partly supported by a Grant-in Aid for Scientific Research 
No. 12640237 from the Ministry of Education, Culture, Sports, 
Science and Technology.

\end{document}